\documentclass[aps,pra,twocolumn,10pt,showpacs]{revtex4}
\usepackage{amsmath}
\usepackage{amsthm}
\usepackage{graphicx}
\usepackage{times}
\usepackage{amssymb}
\usepackage{hyperref}
\usepackage{textcomp}
\usepackage{xcolor}
\usepackage{enumerate}
\usepackage[final]{pdfpages}
\usepackage{multirow}
\usepackage{tabularx}
\usepackage{dcolumn}
\usepackage{subfigure}

\begin{document}

\title{Structural Quantification of Entanglement}
\author{F. Shahandeh}
\email{farid.shahandeh@uni-rostock.de}
\author{J. Sperling}
\author{W. Vogel}
\affiliation{Arbeitsgruppe Theoretische Quantenoptik, Institut f\"ur Physik, Universit\"at Rostock, D-18055 Rostock, Germany}
\date{\today} 

\begin{abstract}
	We introduce an approach which allows a detailed structural and quantitative analysis of multipartite entanglement.
	The sets of states with different structures are convex and nested.
	Hence, they can be distinguished from each other using appropriate measurable witnesses.
	We derive equations for the construction of optimal witnesses and discuss general properties arising from our approach.
	As an example, we formulate witnesses for a 4-cluster state and perform a full quantitative analysis of the entanglement structure in the presence of noise and losses.
	The strength of the method in multimode continuous variable systems is also demonstrated by considering a dephased GHZ-type state.
\end{abstract}


\maketitle

\section{Introduction}

The understanding of the phenomenon entanglement as a fundamental quantum effect grew along with the development of quantum mechanics itself~\cite{EPR35,S35,Nielsen,HorodeckiRev}.
This quantum correlation within compound systems serves as a resource for quantum computation beyond classical limitations, e.g., quantum dense coding~\cite{Bennett}.
It has also been shown that the amount of entanglement of components of a quantum system is of great importance for quantum communication protocols~\cite{Parker,Furusawa}, 
quantum computation~\cite{Eisert-comp}, or quantum metrology~\cite{GZNEO10}.
Therefore, proper methods of classification and quantification of entanglement are essential for applications.

In the case of bipartite quantum systems, a well-established approach for entanglement verification is the witnessing method~\cite{HHH96,HHH01,Terhal1,Lewen1}.
Owing to its simplicity of application, it soon became a popular method for identifying entanglement within experiments~\cite{Eltschka,Huang}. 
Recently, we have studied the construction of optimal witnesses for bipartite systems~\cite{Sperling1}.
Beyond the entanglement detection, an axiomatic approach to entanglement measures has been formulated in different ways, see, e.g.,~\cite{Vedral1,H2001,Plenio,Guhne}.
Therefore, there are many different bipartite entanglement measures, such as relative entropy of entanglement~\cite{Vedral1,Vedral2}, 
geometric measures of entanglement~\cite{Shimony,Wei}, and global robustness~\cite{Vidal,Harrow}.
However, none of these measures are easily implemented in experiments, since in general they require a full quantum-state reconstruction.
In contrast, the bipartite Schmidt number of the state obeys all the conditions of an entanglement measure~\cite{TH00,SBL01}, while it is experimentally accessible through a witnessing approach. 
Moreover, it has been shown that the bipartite Schmidt number is a universal entanglement measure which does not increase under separable operations~\cite{Sperling4}.
Optimal bipartite Schmidt number witnesses have been established in~\cite{Sperling3,Shahandeh}.

The study of multipartite entanglement, however, is a more challenging problem, since for such systems there are many possible ways to subdivide the system into parties.
Multipartite entanglement witnesses render it possible to identify entanglement for any partitioning of the compound system~\cite{Sperling2}. 
In this context, the structuring of the complex states plays an important role~\cite{Huber,Guo,VSK13}.
Even the simplest example of three qubits exhibits nonequivalent GHZ-type and W-type structures of entanglement~\cite{GHZ89,DVC00}. 
In the case of multipartite qubit systems, a graph-theoretic approach allows a more general structural classification~\cite{Raussendorf,RHBM13}.
Moreover, there exist methods for the characterization and possible applications of higher dimensional systems of distinguishable~\cite{AFOV08,LM13} or indistinguishable~\cite{GKM11,KCP14} particles.
Another problem of multipartite entanglement appears when one examines the so-called continuous variable (CV) systems~\cite{vanLoockB,vanLoock1,AdessoR,Shchukin2006,Lian,GK14}.
Due to the complexity of such systems, only a few experimentally accessible quantifiers are known for multipartite CV Gaussian states~\cite{Adesso1}.
However, a multipartite generalization of the Schmidt number has been introduced in~\cite{Eisert}, which is appropriate for both discrete and continuous variable systems.

In the present contribution, we will introduce structural quantifiers of entanglement (SQE) for multipartite quantum systems.
They are based on the witnessing of the multipartite Schmidt number (MSN) which counts the minimal number of global quantum superpositions of product states for different partitions of the compound system under study.
As usual for witnesses, our SQE will be accessible in experiments.
We derive a set of optimization equations for the construction of witnesses for discrete and continuous variable quantum systems.
The method is applied for the full analysis of the SQE for both a $4$-cluster state 
and a multipartite CV GHZ-type state in the presence of noise.

\section{Definition of structural quantifiers}\label{SecII}

Let us study a system with $N$ subsystems which is described through a joint Hilbert space $H{=}\bigotimes_{i\in \boldsymbol I}H_i$.
The index set $\boldsymbol{I}{=}\{1,2,\dots,N\}$ represents the set of the individual subsystems.
Consider a splitting of the Hilbert space $H$ into $n$ parties ($n \leq N$).
We may represent such a specific $n$-partition, $\boldsymbol{P}_n$, of the subsystems by 
a collection of $n$ disjoint nonempty subsets of $\boldsymbol{I}$, $\boldsymbol{P}_n{:=}\{\boldsymbol{I}_1,\dots,\boldsymbol{I}_n\}$,
such that $\bigcup_{i=1}^n \boldsymbol{I}_i=\boldsymbol{I}$.
Now, the Hilbert subspace of the $q$th party is given as $\tilde{H}_q{:=}\bigotimes_{i\in \boldsymbol{I}_q}H_i$.
Any state $|\phi\rangle$ of the system can be decomposed into a sum -- with respect to the $n$-partition $\boldsymbol P_n$ -- as
\begin{equation}
\label{MSD2}
 |\phi \rangle =\sum_{i=1}^r |a_i^{(1)},\dots,a_i^{(n)}\rangle,
\end{equation}
with $r\in\mathbb N\cup\{\infty\}$, $|a_i^{(q)}\rangle\in \tilde H_q$, and the normalization $\langle\phi|\phi\rangle=1$.

Following the approach of Ref.~\cite{Eisert}, Eq.~\eqref{MSD2} represents the optimal decomposition of the pure state $| \phi \rangle$ with respect to the partition $\boldsymbol{P}_n$ when $r$ is minimal.
This is called the multipartite Schmidt decomposition;
$r=r(| \phi \rangle)$ is referred to as the multipartite Schmidt rank, and the product states $|a_i^{(1)},\dots,a_i^{(n)}\rangle$ are the multipartite Schmidt vectors of $| \phi \rangle$.
In contrast to the bipartite case, there is no need for the set of vectors $\{|a_i^{(q)}\rangle\}_{i=1}^r$ in~\eqref{MSD2} to form an orthogonal (or even linearly independent) set of vectors in $\tilde{H}_q$.
That is, one relaxes the biorthonormality condition of standard bipartite Schmidt decomposition~\cite{Nielsen}.
The closure of all pure states having a rank less than or equal to $r$ with respect to the partition $\boldsymbol{P}_n$ defines the set $\boldsymbol{S}_{\boldsymbol{P}_n;r}^{\rm{pure}}$.

One can also extend the above definition to mixed states using a closed convex roof construction. 
The MSN of a mixed state with respect to the partition $\boldsymbol{P}_n$ is given by~\cite{Eisert}:
\begin{align}
	\label{ConvRoof}
	r(\hat\varrho):=\inf_{\boldsymbol{D}(\hat\varrho)}\sup_{k}r(| \phi_k \rangle),
\end{align}
in which $\boldsymbol{D}(\hat\varrho){=}\{p_k, |\phi_k\rangle{:} \hat\varrho{=}\sum_k p_k | \phi_k \rangle \langle \phi_k |\}$ is the set of all ensemble decompositions of 
$\hat\varrho$ and $r(|\phi_k\rangle)$ is the multipartite Schmidt rank with respect to the same partition $\boldsymbol{P}_n$ for all $|\phi_k\rangle$.
It is also clear that the set of MSN-$r$ states, $\boldsymbol{S}_{\boldsymbol{P}_n;r}$, forms a closed convex set,
\begin{equation}
	\boldsymbol{S}_{\boldsymbol{P}_n;r}=\overline{{\rm conv~}\{|\phi\rangle\langle\phi|:|\phi\rangle\in\boldsymbol{S}_{\boldsymbol{P}_n;r}^{\rm pure}\}}.
\end{equation}
Moreover, in analogy to~\cite{Sperling4}, it can be easily shown that MSN cannot increase under all separable operations and thus, it is a universal entanglement measure.

\begin{figure}[h]
  \includegraphics[width=5cm]{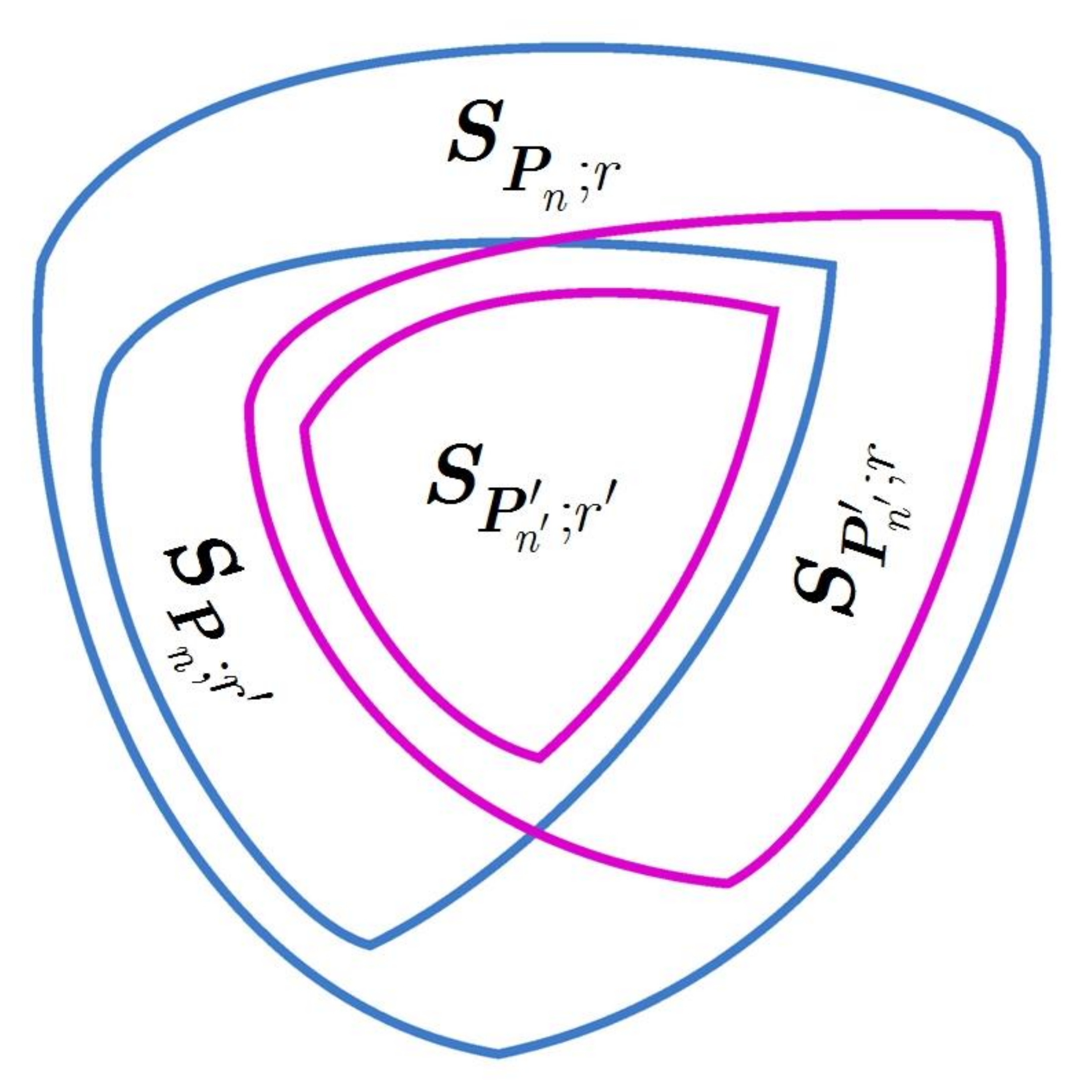}
  \caption{(Color online)
	The schematic representation of the embedded sets of SQE.
	For all refinements $\boldsymbol{P}'_{n'}\preceq\boldsymbol{P}_{n}$ and for all MSNs $r'\leq r$, one has
	$\boldsymbol{S}_{\boldsymbol{P}'_{n'};r'}\subseteq \boldsymbol{S}_{\boldsymbol{P}_n;r}.$
  }\label{EntanglementSetDecomposition}
\end{figure}

Let us discuss how these definitions yield the SQE.
This can be done through a subsequent inclusion of partitions $\boldsymbol P_n$ and MSN-$r$ states, cf. Fig.~\ref{EntanglementSetDecomposition}.
Firstly, from the definition it holds for all $r$ that $\boldsymbol{S}_{\boldsymbol{P}_n;r}\subset \boldsymbol{S}_{\boldsymbol{P}_n;r+1}$.
The number of global superpositions of multipartite product states for a fixed partition $\boldsymbol P_n$ allows the quantification in terms of the MSN $r(\hat\varrho)$ within $\boldsymbol P_n$.
Secondly, a refinement $\boldsymbol{P}'_{n'}=\{\boldsymbol{I}'_1,\dots,\boldsymbol{I}'_{n'}\}$ of the partition $\boldsymbol P_{n}$, $\boldsymbol{P}'_{n'}{\preceq}\boldsymbol{P}_{n}$, fulfilling
$\forall \boldsymbol{I}'_{q'}{\in}\boldsymbol{P}'_{n'}\,\,\exists\boldsymbol{I}_{q}{\in}\boldsymbol{P}_{q}: \boldsymbol{I}'_{q'}{\subseteq}\boldsymbol{I}_{q}$,
is achieved from the original partitioning by further splitting the parties.
This allows a structural study of the entanglement between arbitrary collections of separated parties.
Automatically, we obtain for any refinement that $n'{\geq} n$ and $\boldsymbol{S}_{\boldsymbol{P}'_{n'};r}\subseteq \boldsymbol{S}_{\boldsymbol{P}_n;r}$.
Note that any partition is a refinement of itself, which justifies the semi-ordering $\preceq$.

A determination of both the structural aspect of entanglement in terms of Hilbert space decompositions and the quantitative measure of entanglement by MSN can be combined into the notion of SQE.
In view of SQE, we get a fundamental ordering of multipartite states in the full Hilbert space~\cite{Huber}.
In this way, for all refinements $\boldsymbol{P}'_{n'}\preceq\boldsymbol{P}_{n}$ and for all MSNs $r'\leq r$, it holds that
\begin{align}
	\boldsymbol{S}_{\boldsymbol{P}'_{n'};r'}\subseteq \boldsymbol{S}_{\boldsymbol{P}_n;r}.
\end{align}
This means that the MSN and the refinement of the partitions are compatible properties of multipartite quantum entanglement; see Fig.~\ref{EntanglementSetDecomposition}.
Based on this finding, we will derive a witnessing approach, which allows a simultaneous identification and quantification of the structure of the entangled parties.
We will refer to state $\hat\varrho$ to be entangled with respect to the SQE-$(\boldsymbol{P}_n;r)$ if $\hat\varrho\notin\boldsymbol{S}_{\boldsymbol{P}_n;r}$. 
For example, in Fig.~\ref{EntanglementSetDecomposition} the sets $\boldsymbol{S}_{\boldsymbol{P}_{n;r'}}\setminus\boldsymbol{S}_{\boldsymbol{P}'_{n'};r'}$ and 
$\boldsymbol{S}_{\boldsymbol{P}'_{n';r}}\setminus\boldsymbol{S}_{\boldsymbol{P}'_{n'};r'}$  are both SQE-($\boldsymbol{P}'_{n'};r'$)
entangled.

\section{Optimization procedure}\label{SecIII}

So far, we discussed the construction of nested convex sets with given SQEs.
In this way, determination of SQE of a quantum state requires construction and optimization of appropriate witnesses.
Recently, two approaches for the construction of witnesses -- for multipartite entanglement of any partition~\cite{Sperling2}, and for the bipartite Schmidt number~\cite{Sperling3,Shahandeh} -- has been introduced.
Based on these, we introduce the optimization process of a SQE-$(\boldsymbol{P}_n;r)$ witness operator.
Such a witness, $\hat{\mathcal W}$, is a Hermitian operator such that ${\rm{Tr}}\hat \varrho \hat {\mathcal W}\geqslant 0$ for all states $\hat{\varrho} \in \boldsymbol{S}_{\boldsymbol{P}_n;r}$.
It is also required that ${\rm{Tr}}\hat \varrho \hat {\mathcal W} < 0$ for at least one state $\hat \varrho \notin \boldsymbol{S}_{\boldsymbol{P}_n;r}$.
In general, any witness operator can be written as
\begin{equation}
	\label{witnessform}
	\hat{\mathcal W}=g_r \hat{\mathcal I}{-}\hat{\mathcal L}, 
\end{equation}
with $\hat{\mathcal L}$ -- the so-called {\it test} operator -- being a Hermitian operator.
The witness $\hat{\mathcal W}$ in Eq.~\eqref{witnessform} is optimal by construction,
if $\left\langle \varphi_r  \right|\hat{\mathcal W}\left| \varphi_r  \right\rangle{=}0$ for at least one state $\left| \varphi_r \right\rangle{\in} \boldsymbol{S}_{\boldsymbol{P}_n;r}^{\rm{pure}}$, i.e.,
$g_r:=\left\langle \varphi_r  \right|\hat{\mathcal L}\left| \varphi_r  \right\rangle=\sup_{\left| \phi \right\rangle\in \boldsymbol{S}_{\boldsymbol{P}_n;r}^{\rm{pure}}}
\{\left\langle \phi  \right|\hat{\mathcal L}\left| \phi  \right\rangle\}$.
The SQE-$(\boldsymbol{P}_n,r)$ condition then reads as
\begin{equation}
 \text{Tr}\hat \varrho \hat {\mathcal L} > g_r.
 \label{SQE-cond}
\end{equation}
Thus, the main task is the optimization $\left\langle \phi  \right|\hat{\mathcal L}\left| \phi  \right\rangle \rightarrow g$ constrained to $\langle \phi|\phi \rangle-1=0$.
For this purpose, we use the method of Lagrange's undetermined multiplier, cf. supplemental material~\cite{supp}.
For a specific SQE-$(\boldsymbol{P}_n;r)$, this leads to a set of $n$ simultaneous tensor-operator equations,
\begin{align}
	\label{MSNeq}
	\begin{cases}
	\hat{\mathcal{L}}_{\bar 1}|\vec{a}^{(1)}\rangle=g \hat{\mathcal{I}}_{\bar 1}|\vec{a}^{(1)}\rangle, \\
	\qquad\qquad\vdots\\
	\hat{\mathcal{L}}_{\bar n}|\vec{a}^{(n)}\rangle=g \hat{\mathcal{I}}_{\bar n}|\vec{a}^{(n)}\rangle.
	\end{cases}
\end{align}
In this set of equations, $|\vec{a}^{(q)}\rangle:=(|a_1^{(q)}\rangle,\dots,|a_r^{(q)}\rangle)^{\text{T}}$ are spinors for any party $q\in\{1,\dots,n\}$ and MSN value $r$.
The operator valued matrix $\hat{\mathcal{A}}_{\bar q}$ for $\hat{\mathcal{A}}\in\{\hat{\mathcal{L}},\hat{\mathcal{I}}\}$ is defined through the elements 
$[\hat{\mathcal{A}}_{\bar q}]_{i,j}:=\big(\bigotimes_{p\neq q}\langle a_i^{(p)}|\big)\hat{\mathcal{A}}\big(\bigotimes_{p\neq q}|a_i^{(p)}\rangle\big)$,
which act on the party $\tilde{H}_q$.
Eventually, the optimal value $g_r$ is given by the maximum eigenvalue of the Eqs.~\eqref{MSNeq},
\begin{equation}
\label{Eq:g_r}
	g_r=\max\{g:g\text{ is a SQE-$(\boldsymbol{P}_n;r)$ eigenvalue}\}.
\end{equation}
For the details on the properties regarding SQE eigenvalue equations and their solution, see~\cite{supp}.

\section{SQE for a cluster state}\label{SecIV}

We have solved the set of SQE eigenvalue equations within all possible partitions of a rank-one test operator 
of the form $\hat{\mathcal{L}}=|\psi\rangle\langle\psi|$, with $|\psi\rangle$ being
a $4$-cluster state~\cite{Raussendorf},
\begin{align}
\label{C4-1}
 |\psi\rangle &= \frac{1}{2}\left(|+,0,+,0\rangle + |+,0,-,1\rangle \right.\nonumber \\
 &\qquad\left.+ |-,1,-,0\rangle + |-,1,+,1\rangle\right),
\end{align}
which is a linear combination of product states of four qubit systems.
Here, $|0\rangle$ and $|1\rangle$ are eigenvectors of the Pauli operator $\hat{\sigma}_z$, and $\hat{\sigma}_x|\pm\rangle=\pm|\pm\rangle$.
The results for SQE of such a test operator are given in the Table~\ref{TabbiPart}.
For technical details, please see the supplementary material~\cite{supp}.
\begin{table}
\caption{
	The possible partitions and SQEs of $4$-cluster states are listed.
	This gives the full SQE analysis of the state~\eqref{C4-1}.
	A value $g_r<1$ corresponds to the boundary of entanglement, cf. Eq.~\eqref{SQE-cond}, with respect to the SQE-$(\boldsymbol{P}_n;r)$.
}
\label{TabbiPart}
\begin{ruledtabular}
	\begin{tabular}{ccc}
	Partition                   & $r$ values          & $g_r$ \\ \hline
	$\boldsymbol{P}_{1:2,3,4}$  & $(1,2)$       & $(\frac{1}{2},1)$  \\
	$\boldsymbol{P}_{2:1,3,4}$  & $(1,2)$       & $(\frac{1}{2},1)$  \\
	$\boldsymbol{P}_{3:1,2,4}$  & $(1,2)$       & $(\frac{1}{2},1)$  \\
	$\boldsymbol{P}_{4:1,2,3}$  & $(1,2)$       & $(\frac{1}{2},1)$  \\
	$\boldsymbol{P}_{1,2:3,4}$  & $(1,2)$       & $(\frac{1}{2},1)$  \\
	$\boldsymbol{P}_{1,3:2,4}$  & $(1,2,3,4)$   & $(\frac{1}{4},\frac{1}{2},\frac{3}{4},1)$ \\
	$\boldsymbol{P}_{1,4:2,3}$  & $(1,2,3,4)$   & $(\frac{1}{4},\frac{1}{2},\frac{3}{4},1)$ \\
	$\boldsymbol{P}_{1:2:3,4}$  & $(1,2)$       & $(\frac{1}{2},1)$ \\
	$\boldsymbol{P}_{1:3:2,4}$  & $(1,2,3,4)$   & $(\frac{1}{4},\frac{1}{2},\frac{3}{4},1)$ \\
	$\boldsymbol{P}_{1:4:2,3}$  & $(1,2,3,4)$   & $(\frac{1}{4},\frac{1}{2},\frac{3}{4},1)$ \\
	$\boldsymbol{P}_{2:3:1,4}$  & $(1,2,3,4)$   & $(\frac{1}{4},\frac{1}{2},\frac{3}{4},1)$ \\
	$\boldsymbol{P}_{2:4:1,3}$  & $(1,2,3,4)$   & $(\frac{1}{4},\frac{1}{2},\frac{3}{4},1)$ \\
	$\boldsymbol{P}_{3:4:1,2}$  & $(1,2)$       & $(\frac{1}{2},1)$ \\
	$\boldsymbol{P}_{1:2:3:4}$  & $(1,2,3,4)$   & $(\frac{1}{4},\frac{1}{2},\frac{3}{4},1)$
	\end{tabular}
\end{ruledtabular}
\end{table}
The boundaries for some SQE-$(\boldsymbol{P}_n;r)$ are identical, representing the underlying symmetry of the state.
For example, all SQE eigenvalues of the $4$-partition $\boldsymbol{P}_{1:2:3:4}$ and the $2$-partition $\boldsymbol{P}_{1,3:2,4}$ are the same.
Therefore, whenever a SQE-$(\boldsymbol{P}_{1:2:3:4};r)$ is detected, the state is also entangled with respect to SQE-$(\boldsymbol{P}_{1,3:2,4};r)$.

Now, consider a $4$-cluster state subjected to white noise.
This is a global noise which produces a mixed state,
\begin{equation}
 \hat{\varrho}_{\mu}=\frac{\mu}{16}\bigotimes_{q=1}^{4}\hat{\mathcal{I}}_q + (1-\mu)|\psi\rangle\langle\psi|,
\end{equation}
with $\mu\in[0,1]$ being the amount of white noise.
The expectation value $\text{Tr}\hat{\varrho}_\mu\hat{\mathcal{L}}$
yields the condition $\mu<16(1-g_r)/15$ for certifying a SQE-($\boldsymbol{P}_n,r$), based on the results in Table~\ref{TabbiPart}.
A MSN $r>1$ for any partition, and hence genuine multipartite entanglement, is guaranteed for an amount of white noise of $\mu{<}0.5\bar{3}$.
Partial entanglement persists for $\mu{<}0.8$.

Due to the symmetry, a very interesting situation occurs when there exist local losses -- modelled with beam splitters -- in some channels.
Let us consider losses in the second and fourth parties.
We introduce the transmission and reflection coefficients of the virtual beam splitters as real numbers $t_i$ and $r_i$ ($i=2,4$), satisfying the condition $t_i^2+r_i^2=1$.
After a simple algebra one finds
\begin{equation}
 \text{Tr}\hat{\varrho}_{t_2,t_4}\hat{\mathcal{L}} = \frac{(1+t_1)^2(1+t_4)^2}{16},
\end{equation}
in which $\hat{\varrho}_{t_2,t_4}$ is the mixed state produced as a result of losses.

\begin{figure}[h]%
    \centering
    \subfigure[\label{Loss24}]{\includegraphics[width=6cm]{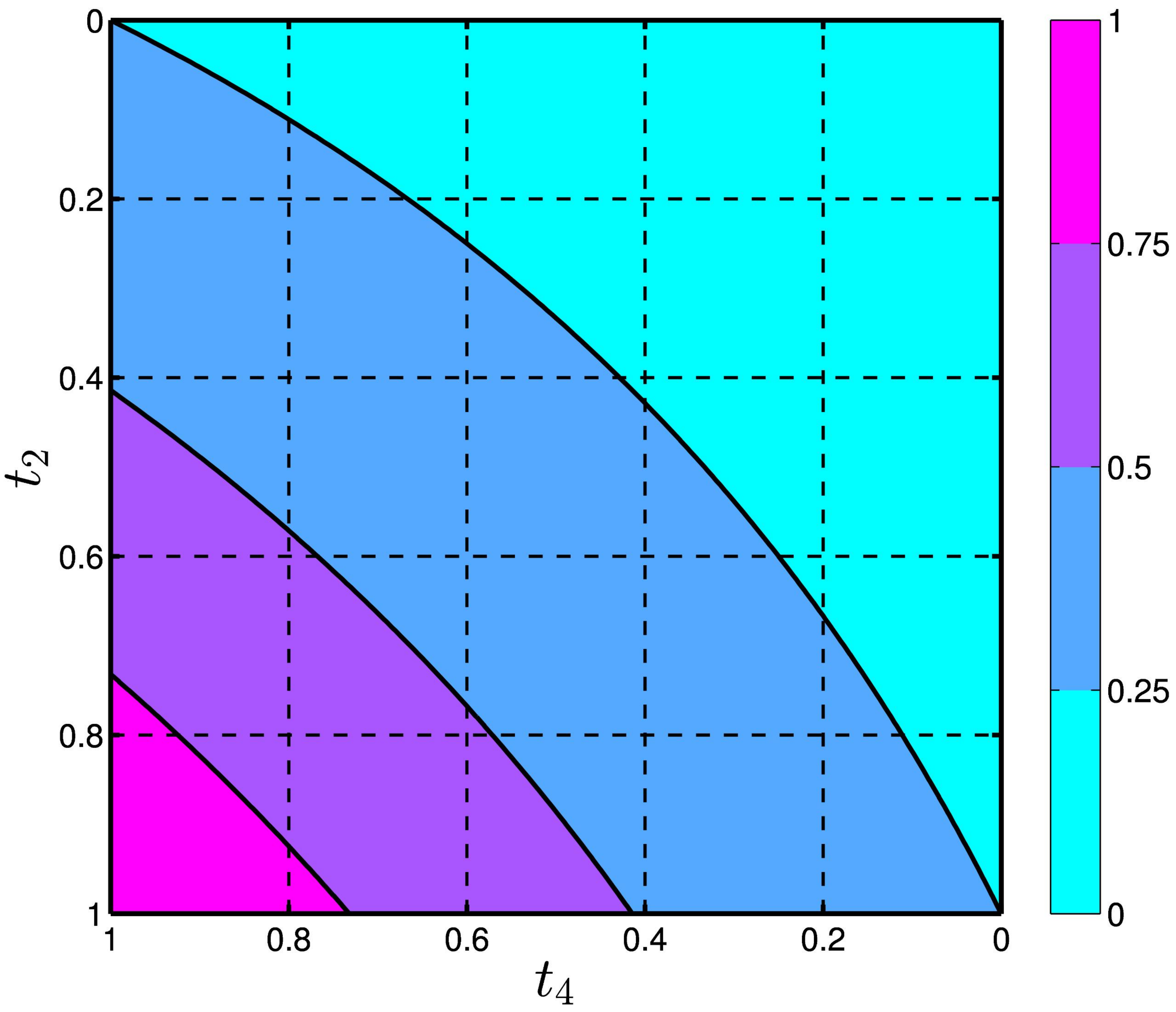} }\\%
    \subfigure[\label{Loss12}]{\includegraphics[width=6cm]{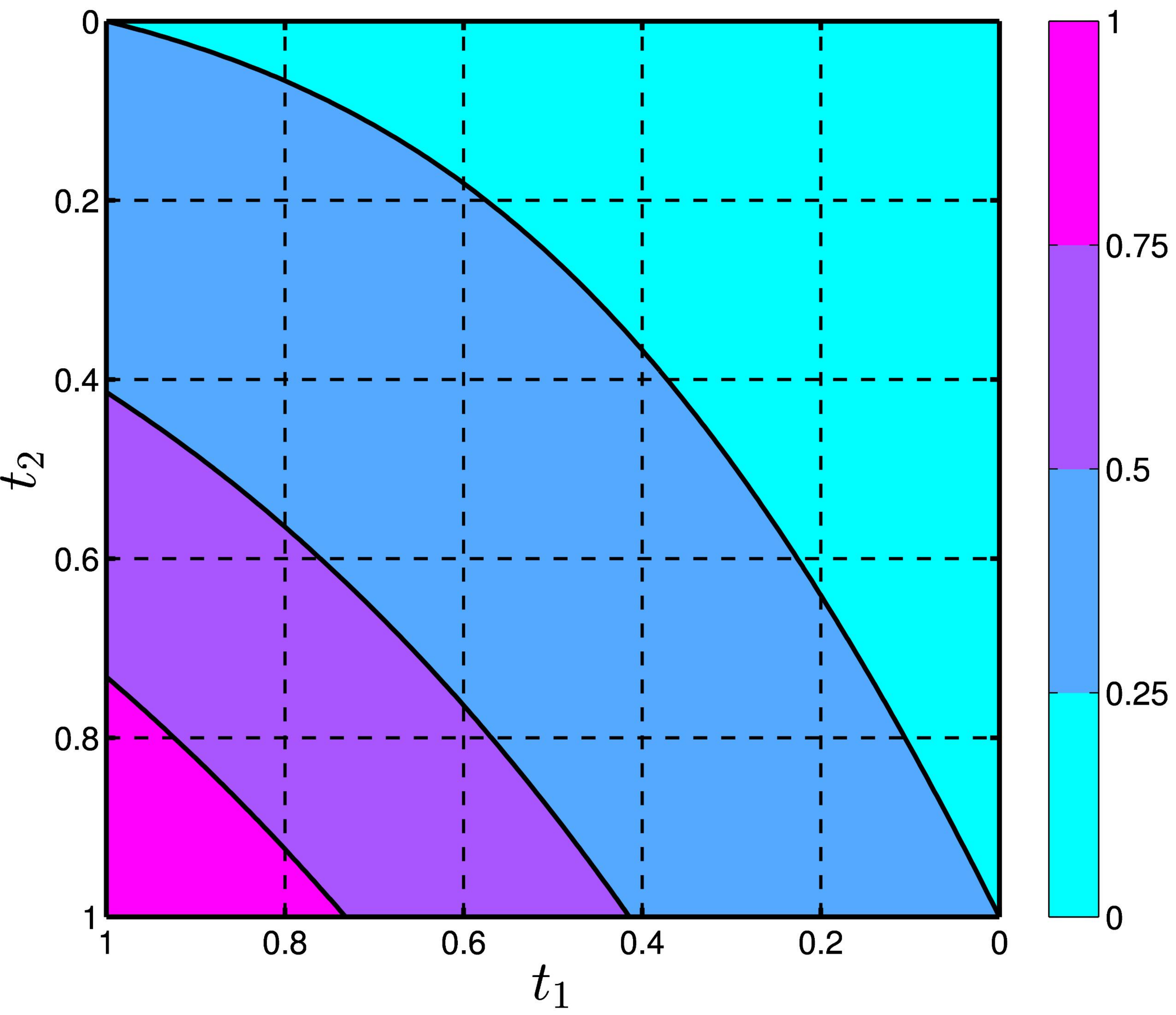} }%
    \caption{(Color online)
	The contour sketches of the expectation value of the test operator, ${\hat {\mathcal L}}$, for a $4$-cluster state undergoing losses from channels
	(a) two and four, and (b) one and two,
	versus the transmission coefficients of each channel.
	The solid lines represent the $g_r$ values within all partitions.
    }%
    \label{fig:example}%
\end{figure}

In Fig.~\ref{Loss24} the contour plot of the expectation value of the witness operator versus two parameters $t_2$ and $t_4$ is depicted.
This again represents the fact that the partial entanglement is very robust.
Also within the regions with expectation values greater than $0.5$, genuine multipartite entanglement can be identified.
A $4$-cluster state is not symmetric with respect to the losses of all channels.
This can be shown through considering losses in the first and second parties.
In such a case, we get
\begin{equation}
\label{TrLoss12}
 \text{Tr}\hat{\varrho}_{t_1,t_2}\hat{\mathcal{L}}=\frac{1}{16}\big[(1+t_1)^2(1+t_2)^2+(1-t_1^2)(1-t_2)^2\big].
\end{equation}
These two are the only channels which create such an asymmetry.
This is due to the special form of the $4$-cluster state which can be written as 
$|\psi\rangle {=} \frac{1}{2}[|+,0\rangle(|+,0\rangle + |-,1\rangle) + |-,1\rangle(|-,0\rangle + |+,1\rangle)]$
with respect to these two parties.
In Fig.~\ref{Loss12} we have represented the contour plot of the expectation value~\eqref{TrLoss12} versus two parameters $t_1$ and $t_2$.
One can simply identify the regions in which an specific amount of SQE is guaranteed.
The asymmetry of the state with respect to losses from the two channels can also be seen.

\section{SQE for a CV GHZ-type state}

We may further analyze a generalized $N$-partite GHZ-type state of the form
\begin{equation}
\label{GHZstate}
 |\psi_{\rm GHZ}\rangle=\sum_{i=0}^{\infty} \lambda_i |i^{(1)},i^{(2)},\dots,i^{(N)}\rangle,
\end{equation}
in which the set $\{|i^{(q)}\rangle\}_{i=1}^{\infty}\subset H_q$ forms an orthogonal basis vector for the Hilbert space of the $q$th subsystem,
and the coefficients $\lambda_i$ are ordered in a descending way.
This can be regarded as a CV $N$-partite GHZ-type state~\cite{vanLoock2}.
Suppose that the state in Eq.~\eqref{GHZstate} suffers from local phase diffusion, described by
$\hat{\mathcal{U}}_q(\theta_q)=\sum_{m=0}^{\infty}\exp(im\theta_q)|m^{(q)}\rangle\langle m^{(q)}|$ and the Gaussian probability distribution 
$p_{\sigma_q}(\theta_q)=(2\pi\sigma_q^2)^{-1/2}\sum_{n\in\mathbb{Z}}\exp{[-(\theta_q+2n\pi)^2/2\sigma_q^2]}$ for each $q=1,\dots,N$.
Bipartite states with this structure have been experimentally produced and also studied in the context of bipartite entanglement~\cite{Kiesel,Mehmet,Sperling5}.
A straight forward calculation yields the dephased state
\begin{align}
\label{GHZPD}
 \hat{\varrho}_{\rm GHZ}&=\sum_{i,j=0}^{\infty}\lambda_i\lambda_j^{\ast}\exp\left[-\frac{\|\vec\sigma\|^2(i-j)^2}{2}\right]\nonumber\\
 &\qquad\quad\times|i^{(1)},\dots,i^{(N)}\rangle\langle j^{(1)},\dots,j^{(N)}|,
\end{align}
in which $\vec{\sigma}:=(\sigma_1,\dots,\sigma_N)$ is the variance vector and $\|\vec \sigma\|^2=\vec \sigma^{\rm T}\vec \sigma$.

For a rank-one test operator of the form $\hat{\mathcal{L}}=|\psi_{\rm GHZ}\rangle\langle\psi_{\rm GHZ}|$, and for any partition $\boldsymbol{P}_n$, 
one finds the MSN-$r$ eigenvalue to be~\cite{supp}
\begin{equation}
\label{grGHZ}
 g_r = \sum_{i=0}^{r-1} |\lambda_i|^2.
\end{equation}
This partition independent structure is due to the fact that reductions of $\hat{\mathcal{L}}$ with respect to any party yield a fully separable operator.
As a direct result of this SQE for generalized $N$-partite GHZ-type states, our witness does not discriminate between different entanglement structures.
Nevertheless, it still identifies the common amount of entanglement for all structures.
According to Eqs.~\eqref{SQE-cond},~\eqref{GHZPD}, and~\eqref{grGHZ}, a common SQE of MSN $r$ is certified if
\begin{align}
\label{GHZcond}
 \sum_{i,j=0}^{\infty}|\lambda_i|^2|\lambda_j|^2\exp\left[-\frac{\|\vec\sigma\|^2(i-j)^2}{2}\right]>\sum_{i=0}^{r-1} |\lambda_i|^2.
\end{align}
Apparently, for $\|\vec\sigma\|\rightarrow 0$ and $\|\vec\sigma\|\rightarrow\infty$ we have entanglement for any $r$ and full separability, respectively.

\begin{figure}[h]
  \includegraphics[width=6cm]{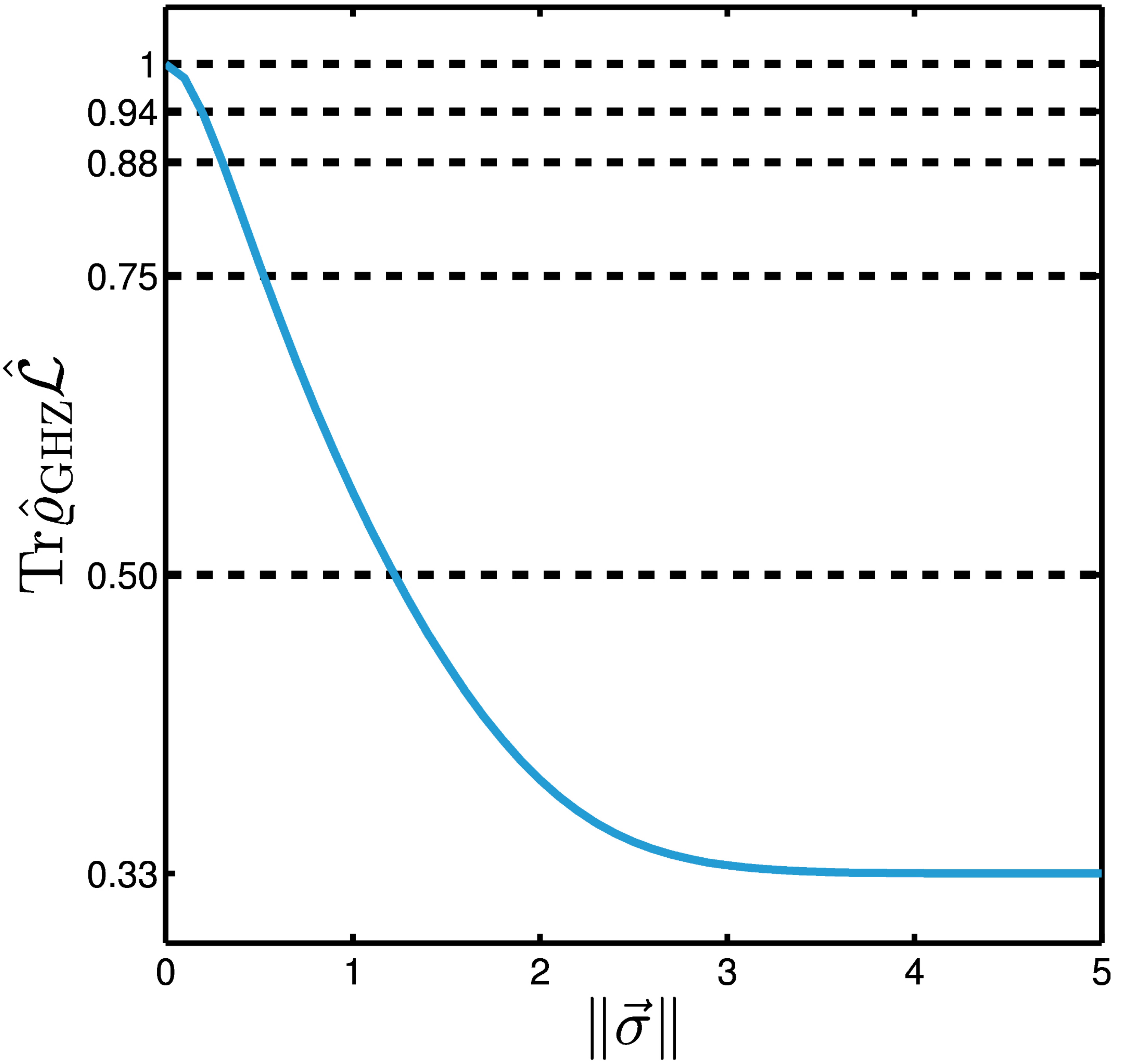}
  \caption{(Color online)
	Semi-logarithmic plot of the expectation value of ${\hat {\mathcal L}}$ (solid line), for a phase diffused $100$-partite CV GHZ-type state
	versus the variance of dephasing.
	From bottom to top, the dashed lines represent the values of $g_r$ for $r=1,2,3,4$ and $\infty$.
  }\label{GHZEx}
\end{figure}

In Fig.~\ref{GHZEx}, we show ${\rm Tr} \hat\varrho_{\rm GHZ}\hat{\mathcal L}$ with $\lambda_i=2^{-(1+i)/2}$ ($i=0,1,\dots$) versus the variance of the dephasing for a $100$-partite state.
When there is no phase diffusion, the inequality~\eqref{GHZcond} holds for any value of $r$.
With increasing phase randomization, the entanglement decreases.
\section{Conclusions}\label{SecV}

In conclusion, we have considered the structural entanglement properties of multipartite systems with respect to a measure called multipartite Schmidt number (MSN).
The sets of states for each partition of the subsystems are shown to be semi-ordered. 
In addition, the sets of states with different MSN values with respect to any partition of the subsystems are convex and possess a nested structure.
The different partions and the corresponding MSNs are combined into the notion of structural quantifiers of entanglement (SQE).
Due to convexity, the SQE can be identified by witnesses, providing a full quantitative and structural entanglement analysis within highly complex quantum systems.
The advantage of SQE is their direct accessibility in experiments.
We have solved the optimization problem of SQE witnesses for the examples of
a lossy $4$-cluster state and a dephased continuous-variable GHZ-type state.

\section*{Acknowledgments}
This work was supported by the Deutsche Forschungsgemeinschaft through SFB~652.



\appendix
\begin{widetext}

\section{SQE Eigenvalue Equation}\label{App:Properties}

\subsection{Derivation of SQE equations}
	Suppose that $\hat{\mathcal{L}}$ is a test operator subjected to a particular partitioning $\boldsymbol{P}_n$ together with a MSN $r$.
	Any pure state, $|\phi\rangle\in\boldsymbol S_{\boldsymbol{P}_n;r}$, can be written as
	\begin{align}
		|\phi\rangle=\sum_{i=1}^r |a_i^{(1)},\dots,a_i^{(n)}\rangle,
	\end{align}
	with $|a_i^{(q)}\rangle\in\tilde{H}_{q}$ for $i=1,\ldots,r$ and $q=1,\ldots,n$.
	The method of Lagrange's undetermined multiplier for the optimization under the normalization constraint reads as
	\begin{align}\label{Eq:ComponentsOpt}
		\forall i\in\{1,\ldots,r\}\quad \forall q\in\{1,\ldots,n\}:\qquad\frac{\partial \langle\phi|\hat{\mathcal L}|\phi\rangle}{\partial \langle a_i^{(q)}|}-g\frac{\partial \langle\phi|\hat{\mathcal I}|\phi\rangle}{\partial \langle a_i^{(q)}|}=0.
	\end{align}
	We have to compute the complex gradients $\partial\langle\phi|\hat{\mathcal A}|\phi\rangle/\partial{\langle a_i^{(q)}|}$ for the expectation value of $\hat{\mathcal A}\in\{\hat{\mathcal L},\hat{\mathcal I}\}$.
	It is useful to define the operator valued matrix $\hat{\mathcal{A}}_{\bar q}$ with elements
	\begin{align}
		\left[\hat{\mathcal{A}}_{\bar q}\right]_{i,j}
		=\langle a_i^{(\bar q)}|\hat{\mathcal{A}}| a_j^{(\bar q)}\rangle,
	\end{align}
	where $|a_i^{(\bar q)}\rangle=| a_i^{(1)},\ldots,a_i^{(q-1)},a_i^{(q+1)},\ldots a_i^{(n)}\rangle\in\bigotimes_{p\ne q}\tilde{H}_{p}$,
	and thus, $[\hat{\mathcal{A}}_{\bar q}]_{i,j}$ is an operator acting on the Hilbert space of the $q$th party, $\tilde{H}_q$.
	We may also define the spinor $|\vec a^{(q)}\rangle=(|a_1^{(q)}\rangle,\dots,|a_r^{(q)}\rangle)^{\rm{T}}$ for any party $q\in\{1,\dots,n\}$.
	Now, the derivatives in Eq.~\eqref{Eq:ComponentsOpt} may be written as
	\begin{align}
		\left(\frac{\partial \langle\phi|\hat{\mathcal A}|\phi\rangle}{\partial \langle a_1^{(q)}|},\dots,\frac{\partial \langle\phi|\hat{\mathcal A}|\phi\rangle}{\partial \langle a_r^{(q)}|}\right)^{\rm{T}}=
		\frac{\partial \langle\vec a^{(q)}|\hat{\mathcal{A}}_{\bar q}|\vec a^{(q)}\rangle}{\partial \langle \vec a^{(q)}|}=\hat{\mathcal{A}}_{\bar q}|\vec a^{(q)}\rangle.
	\end{align}
	Finally, the set of Eqs.~\eqref{Eq:ComponentsOpt} results in the $n$ SQE eigenvalue equations~\eqref{MSNeq},
	\begin{align}
	\label{Eq:SQEeq}
		\hat{\mathcal{L}}_{\bar q}|\vec a^{(q)}\rangle-g\hat{\mathcal{I}}_{\bar q}|\vec a^{(q)}\rangle=0,
		\quad\text{for}\quad q=1,\ldots,n.
	\end{align}
	Multiplying this equation with $\langle\vec a^{(q)}|$ yields $\langle\phi|\hat{\mathcal L}|\phi\rangle=g$, where we have used the normalization $\langle\phi|\hat{\mathcal I}|\phi\rangle=1$.
	Hence, an optimal expectation value of $\hat{\mathcal L}$ in the set $\boldsymbol S_{\boldsymbol{P}_n;r}$ is determined by the multiplier $g$.
	For all refinements $\boldsymbol{P}'_{n'}\preceq\boldsymbol{P}_{n}$ and for all SQEs $r'\leq r$ it holds that $\boldsymbol{S}_{\boldsymbol{P}'_{n'};r'}\subseteq \boldsymbol{S}_{\boldsymbol{P}_n;r}$.
	Thus, $|\varphi_r\rangle\in\boldsymbol{S}_{\boldsymbol{P}'_{n'};r'}$, a solution of the SQE eigenvalue equations~\eqref{Eq:SQEeq} in $\boldsymbol{S}_{\boldsymbol{P}'_{n'};r'}$,
	is automatically a solution for SQE-($\boldsymbol{S}_{\boldsymbol{P}_{n};r}$).

\subsection{Second form of the SQE eigenvalue equations}
	There exist an equivalent form to the set of Eqs.~\eqref{Eq:SQEeq}, called the second form of the SQE eigenvalue equations.
	Consider the action of $\hat{\mathcal L}$ on the SQE eigenvector $|\varphi_r\rangle$,
	\begin{align}\label{Eq:DecAction}
		\hat{\mathcal L}\sum_{i=1}^r|a_i^{(1)},\ldots,a_i^{(n)}\rangle=\tilde g\sum_{i=1}^r|a_i^{(1)},\ldots,a_i^{(n)}\rangle+|\chi\rangle,
	\end{align}
	where the result of $\hat{\mathcal L}|\varphi_r\rangle$ is decomposed into the parallel and orthogonal components, $\tilde g|\varphi_r\rangle$ and $|\chi\rangle\perp|\varphi_r\rangle$, respectively.
	Firstly, we multiply this equation with $\langle\varphi_r|$ and we use $\langle\varphi_r|\varphi_r\rangle=1$ to get $\tilde g=g$.
	Secondly, we multiply~\eqref{Eq:DecAction} with $\langle a_{j}^{(\bar q)}|$ and we find that $\langle a_{j}^{(\bar q)}|\chi\rangle\in\tilde H_q$ has to be the null vector in order to fulfill the first form of the equations.
	Hence, we may define the second form of SQE eigenvalue equations as
	\begin{align}\label{Eq:SecondForm}
		\hat{\mathcal L}|\varphi_r\rangle=g|\varphi_r\rangle+|\chi\rangle,
		\quad\text{with}\quad
		\forall j\in\{1,\ldots,r\}\,
		\forall q\in\{1,\ldots,n\}\,
		\forall |x^{(q)}\rangle \in\tilde H_q:\,
		\quad
		\langle a_{j}^{(\bar q)},x^{(q)}|\chi\rangle=0.
	\end{align}
	In the first form, our equation represent a coupled system of eigenvalue equations,
	whereas in the second form, we get a single but perturbed eigenvalue equation.

\subsection{Local transformations}
	We consider the following transformation of $\hat{\mathcal{L}}$:
	\begin{equation}
		\label{unitrans}
		\hat{\mathcal{L}}^{\prime}:=\left(\bigotimes_{q=1}^{n}\hat{\mathcal{U}}^{(q)}\right)(\xi_1\hat{\mathcal{I}}+\xi_2\hat{\mathcal{L}})\left(\bigotimes_{q=1}^{n}\hat{\mathcal{U}}^{(q)\dag}\right),
	\end{equation}
	in which $\hat{\mathcal{U}}^{(q)}$ is a unitary transformations acting on the $q$th party Hilbert space, $\tilde{H}_q$, and $\xi_{1(2)}$ are nonzero real numbers.
	Given a solution for $\hat{\mathcal{L}}$ to be the SQE eigenvalue $g$ for the SQE eigenvector $|\varphi_r\rangle$, the SQE eigenvector and SQE eigenvalue of the operator $\hat{\mathcal{L}}^{\prime}$ in Eq.~\eqref{unitrans} are given by
	\begin{align}
		|\varphi_r^{\prime}\rangle=\bigotimes_{q=1}^{n}\hat{\mathcal{U}}^{(q)}|\varphi_r\rangle
		\quad\text{and}\quad
		g^{\prime}=\xi_1+\xi_2 g.
	\end{align}
	To prove this claim, we substitute $\hat{\mathcal{L}}^{\prime}$ and $|\varphi_r^{\prime}\rangle$ into the second form of SQE eigenvalue equations.
	We have
	\begin{align}
		\hat{\mathcal{L}}^{\prime}|\varphi_r^{\prime}\rangle &= \bigotimes_{q=1}^{n}\hat{\mathcal{U}}^{(q)}(\xi_1\hat{\mathcal{I}}+\xi_2\hat{\mathcal{L}})|\varphi_r\rangle 
		=(\xi_1 + \xi_2 g)|\varphi_r^{\prime}\rangle + |\chi^{\prime}\rangle,
	\end{align}
	with $|\chi^{\prime}\rangle:=\bigotimes_{q=1}^{n}\hat{\mathcal{U}}^{(q)}|\chi\rangle$.
	Due to unitarity and locality of the transformations $\hat{\mathcal{U}}^{(q)}$, it is clear that $|\chi^{\prime}\rangle$ satisfies the orthogonality condition:
	$\left(\langle a_i^{(\bar q)}|\otimes\langle x^{(q)}|\right)|\chi\rangle=0$ for all $|x^{(q)}\rangle \in \tilde{H}_q$.
	Hence, $|\varphi_r^{\prime}\rangle$ is the true SQE eigenvector of $\hat{\mathcal{L}}^{\prime}$.

	In particular, this relation holds for the largest SQE eigenvalue $g_r'=\xi_1+\xi_2 g_r$ if $\xi_2>0$; in case $\xi_2<0$ one should replace $g_r$ with the minimal SQE eigenvalue of $\hat{\mathcal{L}}$.
	This result implies that one may put test operators into some equivalence classes using transformation~\eqref{unitrans}.
	In this way, if we can solve the SQE eigenvalue equations for a test operator $\hat{\mathcal{L}}$, we have solved the problem for the whole equivalence class elements.

\subsection{Relation to separability in the spinor space}
	In the previous sections, we have occasionally used a spinor representation $|\vec a^{(q)}\rangle\in\tilde H_q\otimes\mathbb C^r$.
	In the following, we are going to represent the set of SQE eigenvalue equations and its solution in a similar way.
	Using the standard orthonormal basis $\{\vec e_{1},\ldots,\vec e_{r}\}$ for $\mathbb C^r$, it is possible to write any element of $\tilde H_q\otimes\mathbb C^r$ as
	$|\vec a^{(q)}\rangle=\sum_{i=1}^r |a_i^{(q)}\rangle\otimes \vec e_i$.
	In this way, any separable vector of the compound spinor space $H\otimes \mathbb C^{nr}=\bigotimes_{q=1}^n\left(\tilde H_q\otimes\mathbb C^r\right)$ is given as
	\begin{align}
		|\vec a^{(1)},\dots,\vec a^{(n)}\rangle=\sum_{i_1=1}^r\dots\sum_{i_n=1}^r |a_{i_1}^{(1)}\rangle\otimes\dots\otimes|a_{i_n}^{(n)}\rangle\otimes \vec e_{i_1}\otimes\dots\otimes \vec e_{i_n}.
	\end{align}
	Now, one can use the vector $\vec s=\sum_{i=1}^r \vec e_i{}^{\otimes n}\in\mathbb C^{nr}$ to relate the SQE $r$ states, $|\varphi_r\rangle=\sum_{i=1}^r |a^{(1)}_i,\ldots,a^{(n)}_i\rangle$, with the separability in the compound spinor space, $|\vec a^{(1)},\dots,\vec a^{(n)}\rangle$, through
	\begin{align}
		\langle \vec a^{(1)},\dots,\vec a^{(n)}|\hat{\mathcal L}\otimes \vec s\vec s^{\,\dagger}|\vec a^{(1)},\dots,\vec a^{(n)}\rangle=\langle \varphi_r|\hat{\mathcal L}|\varphi_r\rangle,
	\end{align}
	including the normalization $\langle \vec a^{(1)},\dots,\vec a^{(n)}|\hat{\mathcal I}\otimes \vec s\vec s^{\,\dagger}|\vec a^{(1)},\dots,\vec a^{(n)}\rangle=\langle \varphi_r|\varphi_r\rangle=1$.
	Hence, the MSN $r$ detection in $\bigotimes_{q=1}^n\tilde H_q$ can be mapped to the separability problem in the spinor space $\bigotimes_{q=1}^n\left(\tilde H_q\otimes\mathbb C^r\right)$.
	Note that a possible way to construct entanglement witnesses can be found in Ref.~\cite{Sperling2}.

\subsection{Cascade structure of SQE equations} \label{CasStr}
	In Ref.~\cite{Sperling2}, the so-called cascade structure of the multipartite separability eigenvalue equations has been introduced to reduce the number of parties contributing in a test operator.
	A very useful extension of the approach can be obtained for SQE eigenvalue equations as well.
	Consider an operator $\hat{\mathcal{L}}$ of the form
	\begin{equation}
		\hat{\mathcal{L}}=\sum_{m=1}^{\infty} |\psi_m\rangle \langle\psi_m|,
	\end{equation}
	in which each $|\psi_m\rangle \in H$ has a bipartite Schmidt decomposition as	$|\psi_m\rangle=\sum\limits_{j}\lambda_{m,j} |\psi_{m,j}^{(\bar n)}\rangle |\psi_{m,j}^{(n)}\rangle$.
	Let us define the set of operators,
	$\hat{\mathcal{L}}_{m,m^{\prime}}:=|\psi_m\rangle \langle\psi_{m^{\prime}}|$, with $m,m^{\prime}\in\mathbb N$.
	Now, according to the $q$th equation ($q\neq n$) of the SQE eigenvalue equations, the $(k,l)$ component of the tensor-operator $\hat{\mathcal{L}}_{\bar q}$ reads as
	\begin{align}
		\label{klLq}
		\left[\hat{\mathcal{L}}_{\bar q}\right]_{k,l}=\sum\limits_{m=1}^{\infty}\sum\limits_{j} \lambda_{m,j} \langle a_{k}^{(\bar q,\bar n)}|\psi_{m,j}^{(\bar n)}\rangle \langle a_{k}^{(n)}|\psi_{m,j}^{(n)}\rangle
		\langle\psi_{m}|a_{l}^{(\bar q)}\rangle,
	\end{align}
	where we have only used the Schmidt decomposition of the ket vectors.
	By substituting $\langle a_{k}^{(n)}|$ from the $k$th component for the $n$th party of the SQE eigenvector into Eq.~\eqref{klLq}, rearranging the terms and summing over the index $j$, we obtain
	\begin{align}
		\hat{\mathcal{L}}_{\bar q}|\vec{a}^{(q)}\rangle = 
		[\hat{\mathcal{I}}_{\bar n}^{-1}]^{\ast} \circ \sum_{m,m^{\prime}=1}^\infty \sigma_{m,m^{\prime}}  \hat{\mathcal{L}}_{m,m^{\prime};\bar q}^{(\bar n)} |\vec{a}^{(q)}\rangle,
	\end{align}
	in which $\circ$ is the Hadamard or element-wise matrix product,
	\begin{equation}
		\label{Sigma}
		\hat{\mathcal{L}}_{m,m^{\prime}}^{(\bar n)}{:=}{\rm Tr}_n \hat{\mathcal{L}}_{m,m^{\prime}},\quad \text{and}\quad
		\sigma_{m,m^{\prime}}:=\frac{1}{g}\langle\vec{a}^{(q)}|\hat{\mathcal{L}}_{m,m^{\prime};\bar q}|\vec{a}^{(q)}\rangle.
	\end{equation}
	From $\hat{\mathcal L}=\hat{\mathcal L}^\dagger$ follows $\hat{\mathcal{L}}_{m,m^{\prime}}^{\dag} = \hat{\mathcal{L}}_{m^{\prime},m}$ and $\sigma_{m,m^{\prime}}^{\ast}=\sigma_{m^{\prime},m}$.
	The definition of $\sigma_{m,m'}$ additionally implies $\sum_{m=1}^\infty \sigma_{m,m} = 1$.
	The SQE equations now can be written in the form,
	\begin{equation}
	\label{redMSNMix}
		[\hat{\mathcal{I}}_{\bar n}^{-1}]^{\ast} \circ \sum_{m,m^{\prime}=1}^\infty \sigma_{m,m^{\prime}}  \hat{\mathcal{L}}_{m,m^{\prime};\bar q}^{(\bar n)} |\vec{a}^{(q)}\rangle
		= g \hat{\mathcal{I}}_{n} \circ \hat{\mathcal{I}}_{\bar n,\bar q}|\vec{a}^{(q)}\rangle.
	\end{equation}
	A relatively simple case to handle this equation is that of $r{=}1$. 
	That is to say, the multipartite separability eigenvalue equations are relatively easy to solve for a mixture of projection operators.
	Because, during the procedure above, the rank of the test operator does not increase.
	
\section{Solution Strategies}\label{App:Solutions}

\subsection{Identifying the solution by bipartite Schmidt decompositions}
	We may further study the second form of the SQE eigenvalue problem~\eqref{Eq:SecondForm}.
	For any party $q\in\{1,\ldots,n\}$, one can perform a bipartite Schmidt decomposition of the eigenvector,
	\begin{equation}
	 |\varphi_r\rangle=\hat{\mathcal U}^{(\bar q)}\otimes\hat{\mathcal U}^{(q)}\sum_{i=1}^{r_q} \lambda_i|i^{(\bar q)},i^{(q)}\rangle,
	 \quad \text{with } r_q\leq r,
	\end{equation}
	in which $\{|i^{(\bar q)}\}_{i=1}^{r_q}$ and $\{|i^{(q)}\}_{i=1}^{r_q}$ form orthonormal basis for $\bigotimes_{p\ne q}\tilde{H}_p$ and $\tilde{H}_q$.
	Moreover, $\hat{\mathcal U}^{(\bar q)}$ and $\hat{\mathcal U}^{(q)}$ are unitary transformations acting on the corresponding Hilbert spaces.
	From Eq.~\eqref{Eq:SecondForm} for the perturbation term $|\chi\rangle$ it holds that $\langle x^{(q)}|\chi\rangle=0$ for any $| x^{(q)}\rangle\in \tilde{H}_q$,
	and thus, $\langle i^{(\bar q)},x^{(q)}| \left(\hat{\mathcal U}^{(\bar q)}\otimes\hat{\mathcal U}^{(q)}\right)^\dagger|\chi\rangle=0$ for all $i=1,\ldots,r_q$.
	Therefore, we can expand $|\chi\rangle$ as
	\begin{align}
		|\chi\rangle=\hat{\mathcal U}^{(\bar q)}\otimes\hat{\mathcal U}^{(q)} \sum_{i\notin\{1,\ldots,r_q\}} |i^{(\bar q)},\chi^{(q)}_i\rangle,
	\end{align}
	with $|\chi^{(q)}_i\rangle\in\tilde H_q$.
	We may formulate this result as follow:
	$|\varphi_r\rangle$ is an SQE eigenvector of $\hat{\mathcal L}$, iff
	\begin{align}\label{eq:decompSQEvector}
		\forall q\in\{1,\ldots,n\}\,
		\exists \hat{\mathcal U}^{(\bar q)}\text{ unitary: }\,
		\hat{\mathcal U}^{(\bar q)}\otimes \hat{\mathcal I}^{(q)}|\varphi_r\rangle=\sum_{i=1}^{r_q}|i^{(\bar q)},\phi_i^{(q)}\rangle
		\quad\text{and}\quad
		\hat{\mathcal U}^{(\bar q)}\otimes \hat{\mathcal I}^{(q)}|\chi\rangle=\sum_{i\notin\{1,\ldots,r_q\}}|i^{(\bar q)},\chi_i^{(q)}\rangle,
	\end{align}
	with $|\chi\rangle=(\hat{\mathcal L}-g\hat{\mathcal I})|\phi_r\rangle$, orthonormal $\{|i^{(\bar q)}\rangle\}_{i=1}^{r_q}$, and $|\phi^{(q)}_i\rangle,|\chi^{(q)}_i\rangle\in\tilde H_q$.
	This means that, for any choice of $q$, the $\bar q$ party components of the eigenvector $|\varphi_r\rangle$ and the perturbation $|\chi\rangle$ possess a common orthogonal expansion basis.

\subsection{Solutions based on unextendible/orthonormal product basis}
	Consider projection operators $\hat{\mathcal{L}}=|\psi\rangle\langle\psi|$ in continuous variables -- which can be easily restricted to finite dimensional subspaces.
	The second form of the SQE eigenvalue equations~\eqref{Eq:SecondForm} implies,
	\begin{align}
	\label{Eq:DecPsi}
		\hat{\mathcal{L}}|\varphi_r\rangle=g|\varphi_r\rangle+|\chi\rangle
		\quad\Rightarrow\quad
		|\psi\rangle=\gamma^\ast|\varphi_r\rangle+\frac{1}{\gamma}|\chi\rangle,
		\quad\text{with}\quad\gamma=\langle\psi|\varphi_r\rangle.
	\end{align}
	This can be studied in connection with the finding in Eq.~\eqref{eq:decompSQEvector},
	where we restrict our consideration to non-vanishing eigenvalues $g=|\gamma|^2>0$ to find the maximal value $g_r=\sup\{g\}$ for the positive semidefinite operator $\hat{\mathcal{L}}$.

	Let us assume that the vector $|\psi\rangle$ has the structure
	\begin{align}\label{Eq:UPBstatedec}
		|\psi\rangle=\sum_{k=1}^\infty \kappa_k |b_k^{(1)},\ldots,b_k^{(n)}\rangle,
		\quad\text{with}\quad
		\forall q\in\{1,\ldots,n\}: \langle b_k^{(\bar q)}|b_{k'}^{(\bar q)}\rangle=\delta_{k,k'},
	\end{align}
	and ordered coefficients $|\kappa_1|\geq|\kappa_2|\geq\cdots$.
	Hence, $\{|b_k^{(\bar q)}\rangle\}_k$ defines a (multipartite) unextendible product -- sometimes even orthonormal -- basis for each $\bar q$~\cite{UPB}.
	Since, $|\psi\rangle$ is already in the form of Eq.~\eqref{Eq:DecPsi}, we get the SQE eigenvalue and SQE vector as
	\begin{align}\label{Eq:UPBstatedecSolution}
		g=\sum_{k=1}^r |\kappa_k|^2
		\quad\text{for}\quad
		|\varphi_r\rangle=\left(\sum_{k=1}^r |\kappa_k|^2\right)^{-1/2} \sum_{k=1}^\infty \kappa_k |b_k^{(1)},\ldots,b_k^{(n)}\rangle.
	\end{align}
	However, due to the fact that the decomposition in Eq.~\eqref{Eq:UPBstatedec} is not unique, one should consider a maximization over all possible such a decompositions:
	\begin{equation}
	 g_r=\sup_{\boldsymbol D(|\psi\rangle)}\left\{g:g=\sum_{k=1}^r |\kappa_k|^2\right\},
	\end{equation}
	where $\boldsymbol D(|\psi\rangle)$ is the set of all decompositions of $|\psi\rangle$ in the form~\eqref{Eq:UPBstatedec}.
	The simplest example of such projector $\hat{\mathcal{L}}=|\psi\rangle\langle\psi|$, is given by a generalized $N$-partite GHZ-type state,
	\begin{equation}
		|\psi\rangle=\sum_{i=1}^{\infty} \lambda_i |i^{(1)},i^{(2)},\dots,i^{(N)}\rangle,
	\end{equation}
	in which the set $\{\bigotimes_{p\in\boldsymbol I_q}|i^{(p)}\rangle\}_{i=1}^{\infty}\subset \tilde{H}_q$ forms even an orthogonal basis for the Hilbert space of the $q$th subsystem.

	More generally, we may study an operator $\hat{\mathcal{L}}$ as
	\begin{align}\label{eq:UPBop}
		\hat{\mathcal{L}}=\sum_{k,l=1}^\infty L_{k,l} |b_k^{(1)},\ldots,b_k^{(n)}\rangle\langle b_l^{(1)},\ldots,b_l^{(n)}|,
	\end{align}
	with unextendible/orthonormal product basis $\langle b_k^{(\bar q)}|b_{k'}^{(\bar q)}\rangle=\delta_{k,k'}$ for $q=1,\ldots,n$.
	Let us additionally assume that $[L_{k,l}]_{k,l}$ is a positive semidefinite (infinite dimensional) coefficient matrix.
	Hence, we have a mapped state of the form
	$\hat{\mathcal{L}}|\varphi_r\rangle=\sum_k \kappa_k |b_k^{(1)},\ldots,b_k^{(n)}\rangle$ for any $|\varphi_r\rangle$,
	which is in the form of Eq.~\eqref{eq:decompSQEvector}.
	Thus, the solutions -- up to permutations of the indices -- are given by
	\begin{align}
		|\varphi_r\rangle=\sum_{k=1}^r \lambda_k |b_k^{(1)},\ldots,b_k^{(n)}\rangle,
		\quad\text{with}\quad
		\sum_{l=1}^r L_{k,l}\lambda_l=g\lambda_k.
	\end{align}
	This means that $[\lambda_l]_l$ is an eigenvector with the eigenvalue $g$ of the ordinary eigenvalue problem of an $r\times r$ sub-matrix of $[L_{k,l}]_{k,l}$.

\subsection{Partially separable operators}
	Further on, we may assume an operator -- exhibiting a partial spectral decomposition -- of the form
	\begin{align}\label{Eq:PartSepOp}
		\hat{\mathcal{L}}=\sum_{i=1}^\infty \hat{\mathcal{K}}_{i}\otimes |i^{(n)}\rangle\langle i^{(n)}|,
	\end{align}
	with orthonormal $\{|i^{(n)}\rangle\}_{i=1}^\infty$ (resolving the identity $\hat{\mathcal{I}}^{(n)}=\sum_{i=1}^\infty |i^{(n)}\rangle\langle i^{(n)}|$), and $\hat{\mathcal{K}}_{i}$ being arbitrary positive semidefinite operators acting on $\bigotimes_{j=1}^{n-1}\tilde H_j$.
	The expectation value of $|\varphi_r\rangle$ in such a case can be written as
	\begin{align}
		\langle\varphi_r|\hat{\mathcal{L}}|\varphi_r\rangle=\sum_{i=1}^\infty \frac{\langle\phi_{r,i}|\hat{\mathcal{K}}_i|\phi_{r,i}\rangle}{\langle\phi_{r,i}|\phi_{r,i}\rangle}\langle\phi_{r,i}|\phi_{r,i}\rangle,
	\end{align}
	with $|\phi_{r,i}\rangle=\sum_{k=1}^r \langle i^{(n)}|a_{k}^{(n)}\rangle|a_k^{(1)},\ldots,a^{(n-1)}_k\rangle$ and the normalization $\sum_{i=1}^\infty\langle\phi_{r,i}|\phi_{r,i}\rangle=1$.
	In this form, we find that the expectation value of $\hat{\mathcal{L}}$ is a convex combination of expectation values of the family of non-negative operators $\{\hat{\mathcal{K}}_i\}_{i}$.
	This is maximized by $\max_{i}\sup\{\langle\tilde \phi_{r}|\hat{\mathcal{K}}_i|\tilde \phi_{r}\rangle:|\tilde \phi_r\rangle\in\boldsymbol S_{\boldsymbol{P'}_{n-1};r}\}$, 
	corresponding to the state $|\varphi_{r,i}^{\prime}\rangle$ in $\boldsymbol{P'}_{n-1}=\boldsymbol{P}_{n}\setminus\boldsymbol{I}_{n}$.
	Subsequently, the maximal SQE eigenvalue is found for a state of the form
	\begin{align}
	\label{Eq:MaxSol}
		|\varphi_{r}\rangle=|\varphi_{r,i}^{\prime}\rangle\otimes |i^{(n)}\rangle,
		\quad\text{with}\quad
		|\varphi'_{r,i}\rangle\in\boldsymbol S_{\boldsymbol{P}_{n-1}^{\prime};r}
		\quad\text{for some}\quad i\in\{1,2,\ldots\},
	\end{align}
	which reduced the problem to one less subsystem.
	This finding is consistent, if we consider the physically intuitive fact that classical mixing of two parties cannot increase the entanglement, cf. Eq.~\eqref{Eq:PartSepOp}.
	
	Now, Let us state a direct corollary of the statement above.
	Consider an operator of the form 
	\begin{equation}
	\label{Eq:PartSepRed}
		\hat{\mathcal{L}}^{(\bar n)}=\sum_{i=1}^\infty |\psi_{i}\rangle\langle\psi_{i}|\otimes |i^{(n-1)}\rangle\langle i^{(n-1)}|,
	\end{equation}
	with $\{|i^{(n-1)}\rangle\}_{i=1}^{\infty}$ being a set of orthonormal vectors in $\tilde{H}_{n-1}$.
	Suppose that $|\varphi_{r,i}^{\prime}\rangle$ is the SQE-$r$ solution of the projection $|\psi_{i}\rangle\langle\psi_{i}|$ corresponding to SQE-$r$ eigenvalue $g_{r,i}$ 
	within some $(n-2)$-partition $\boldsymbol{P}_{n-2}^{\prime}=\boldsymbol{P}_n\setminus\{\boldsymbol{I}_n,\boldsymbol{I}_{n{-}1}\}$.
	This operator can be purified to a $n$-partite test operator $\hat{\mathcal{L}}$, using a set of orthonormal vectors $\{|i^{(n)}\rangle\}_{i=1}^{\infty}$ being elements of the Hilbert space $\tilde{H}_n$, as
	\begin{equation}
	\label{PureLForm}
		\hat{\mathcal{L}}=|\psi\rangle\langle\psi|, 
		\quad\text{with}\quad
		|\psi\rangle=\sum_{i=1}^{\infty} |i^{(n)},i^{(n-1)},\psi_{i}\rangle,
		\end{equation}
	The SQE-$r$ solution of $\hat{\mathcal{L}}$ within the $n$-partition $\boldsymbol{P}_{n}$ can be constructed using the following procedure:
	\begin{enumerate}[(i)]
		\item Consider a $K$-partition of the integer $r$ -- such that $r=\sum_{i=1}^K r_{i}$ and $1\leq K\leq r$ -- and denote the set of all such $K$-partitions as $\boldsymbol{K}$.
		Find the corresponding SQE-$r_i$ eigenvectors of $\hat{\mathcal{L}}^{(\bar n)}$ for all elements in $\boldsymbol K$, 
		namely $|\varphi_{r_i,i}^{\prime}\rangle\otimes|i^{(n-1)}\rangle$, as given in Eq.~\eqref{Eq:MaxSol}.
		Note that the index $i$ cannot be repeated.
		\item Write the SQE-$r$ solution, for all $K$-partitions in $\boldsymbol K$, in the form
		$|\tilde{\varphi}_{r}\rangle = \frac{1}{\sqrt{g}}\sum_{i=1}^{K} \sqrt{g_{r_i}}|a_i^{(n)},a_i^{(n-1)},\varphi_{r_i,i}^{\prime}\rangle$.
		\item Considering the relation~\eqref{Eq:DecPsi} gives the corresponding SQE-$r$ eigenvalue as
		\begin{equation}
		\label{MixEigenval}
			g_r=\max_{1\leq K\leq r}\max_{\boldsymbol K}\left\{\langle\tilde{\varphi}_{r}|\hat{\mathcal{L}}|\tilde{\varphi}_{r}\rangle=\sum_{i=1}^K g_{r_i}\right\}.
		\end{equation}
		\end{enumerate}
	The proof is as follows.
	It is easy to check that the outcome of the construction process above is a solution of the SQE-$r$ eigenvalue problem for the test operator in Eq.~\eqref{PureLForm}.
	It is also clear that within $\boldsymbol{I}_n$ and $\boldsymbol{I}_{n-1}$ the sets $\{|i^{(n)}\rangle\}_{i=1}^{\infty}$ and $\{|i^{(n-1)}\rangle\}_{i=1}^{\infty}$ are optimal, 
	because the operator $\hat{\mathcal{L}}$ is partially diagonalized with respect to these parties. Therefore, decomposition~\eqref{Eq:DecPsi} for $|\psi\rangle$ must contain vector components from
	these two sets for $\boldsymbol{I}_n$ and $\boldsymbol{I}_{n-1}$ parties. Now, from Eq.~\eqref{Eq:MaxSol}, if we choose the solution to contain a vector from 
	$\{|i^{(n-1)}\rangle\}_{i=1}^{\infty}$, it should be accompanied with a SQE-$r_i$ solution for $|\psi_{i}\rangle\langle\psi_{i}|$.
	
	The procedure given above can be applied to an operator of the form
	\begin{equation}
	\label{Eq:PartSepRedOrth}
		\hat{\mathcal{L}}^{(\bar n)}=\sum_{i=1}^{\infty} |\psi_{i}\rangle\langle\psi_{i}|,
	\end{equation}
	if all the $|\psi_{i}\rangle$ vectors are orthogonal and we need a SQE-$r$ eigenvector of $\hat{\mathcal{L}}$ with $r\geq \max\{r(|\psi_{i}\rangle)\}$.
	The claim is as before, except that when $r<\max\{r(|\psi_{i}\rangle)\}$, one cannot state that the solution should contain vectors from $\boldsymbol{I}_n$ and the set $\{|i^{(n)}\rangle\}_{i=1}^{\infty}$.
	However, due to the optimality of the set $\{|i^{(n)}\rangle\}_{i=1}^{\infty}$ for the purification procedure, when $r\geq \max\{r(|\psi_{i}\rangle)\}$, 
	then, the solution must contain at least one of the vectors from this set.
\section{$4$-Cluster state as a test operator}\label{App:ClusterState}
	In the following, we are about to solve the set of SQE eigenvalue equations for the test operator of the form $\hat{\mathcal{L}}=|\psi\rangle\langle\psi|$, with $|\psi\rangle$ being a $4$-cluster state,
	\begin{align}
		|\psi\rangle &= \frac{1}{2}\left(|+,0,+,0\rangle + |+,0,-,1\rangle + |-,1,-,0\rangle + |-,1,+,1\rangle\right),
	\end{align}
	which is a linear combination of product states of four qubit systems.
	Here, $|0\rangle$ and $|1\rangle$ are eigenvectors of the Pauli operator $\hat{\sigma}_z$ and $\hat{\sigma}_x|\pm\rangle=\pm1|\pm\rangle$.
	First, we define the following partitions:
	\begin{align}
		2\text{-partitioning: }\quad
		&\boldsymbol{P}_{1:2,3,4}:=\{\{1\},\{2,3,4\}\}\cong\boldsymbol{P}_{2:1,3,4}\cong\boldsymbol{P}_{3:1,2,4}\cong\boldsymbol{P}_{4:1,2,3},\\
		&\boldsymbol{P}_{1,2:3,4}:=\{\{1,2\},\{3,4\}\}\\
		&\boldsymbol{P}_{1,3:2,4}:=\{\{1,3\},\{2,4\}\}\cong\boldsymbol{P}_{1,4:2,3},\\
		3\text{-partitioning: }\quad
		&\boldsymbol{P}_{1:2:3,4}:=\{\{1\},\{2\},\{3,4\}\}\\
		&\boldsymbol{P}_{1:3:2,4}:=\{\{1\},\{3\},\{2,4\}\}\cong\boldsymbol{P}_{1:4:2,3}\cong\boldsymbol{P}_{2:3:1,4}\cong\boldsymbol{P}_{2:4:1,3},\\
		&\boldsymbol{P}_{3:4:1,2}:=\{\{3\},\{4\},\{1,2\}\},\\
		4\text{-partitioning: }\quad
		&\boldsymbol{P}_{1:2:3:4}:=\{\{1\},\{2\},\{3\},\{4\}\},
	\end{align}
	where the symbol ''$\cong$'' denotes an equivalent partition -- possibly by local flip operations $\hat\sigma_x$ or Hadamard gates $|+\rangle\langle 0|+|-\rangle\langle 1|$ -- for the given symmetries of the state.
	Note that in the trivial one-partition, $\boldsymbol{P}_{1,2,3,4}=\{\{1,2,3,4\}\}$, the SQE eigenvalue equation correspond to the standard eigenvalue problem having the eigenvalue $g_1=1$.

\subsection{$2$-Partitionings}
	Any $2$-partition problem is equivalent to the bipartite Schmidt number problem, which can be handled according to the approach in Ref.~\cite{Sperling3}.
	However, we may use the approach in~\eqref{eq:decompSQEvector}.
	In particular, the bipartition $\boldsymbol{P}_{1:2,3,4}$ refers to the possibility of considering the Schmidt decomposition between the Hilbert space $\tilde{H}_1=H_1$ and the Hilbert space $\tilde{H}_2=H_2\otimes H_3\otimes H_4$,
	\begin{align}
		|\psi\rangle &
		=\frac{1}{\sqrt 2}|+\rangle\otimes \frac{|0,+,0\rangle + |0,-,1\rangle}{\sqrt 2}
		+\frac{1}{\sqrt 2}|-\rangle\otimes\frac{|1,-,0\rangle + |1,+,1\rangle}{\sqrt 2},
	\end{align}
	allowing a rank $r\leq2$ with locally orthonormal vectors.
	This results in $g_1=1/2$ and $g_2=1$ for $|\varphi_1\rangle\in\{|+\rangle\otimes(|0,+,0\rangle + |0,-,1\rangle)/\sqrt 2,|-\rangle\otimes(|1,-,0\rangle + |1,+,1\rangle)/\sqrt 2\}$ and $|\varphi_2\rangle=|\psi\rangle$.
	Analogously, we have for $\boldsymbol{P}_{1,2:3,4}$
	\begin{align}
		|\psi\rangle &
		=\frac{1}{\sqrt 2} |+,0\rangle\otimes \frac{|+,0\rangle + |-,1\rangle}{\sqrt 2}
		+\frac{1}{\sqrt 2} |-,1\rangle\otimes\frac{|-,0\rangle + |+,1\rangle}{\sqrt 2},
	\end{align}
	for a rank $r\leq2$, with $g_1=1/2$ and $g_2=1$,
	and for $\boldsymbol{P}_{1,3:2,4}$,
	\begin{align}
		|\psi\rangle &
		=\frac{1}{2} |+,+\rangle\otimes |0,0\rangle
		+\frac{1}{2} |+,-\rangle\otimes |0,1\rangle
		+\frac{1}{2} |-,+\rangle\otimes |1,1\rangle
		+\frac{1}{2} |-,-\rangle\otimes |1,0\rangle,
	\end{align}
	we have $r\leq4$, with $g_1=1/4$, $g_2=1/2$, $g_3=3/4$, and $g_4=1$.

\subsection{$3$- and $4$-Partitioning}
	For the partitioning $\boldsymbol P_{1:2:3,4}$ we can decompose the state as
	\begin{align}
		|\psi\rangle &
		=\frac{1}{\sqrt 2} |+\rangle\otimes |0\rangle\otimes \frac{|+,0\rangle+|-,1\rangle}{\sqrt 2}
		+\frac{1}{\sqrt 2} |-\rangle\otimes |1\rangle\otimes \frac{|-,0\rangle+|+,1\rangle}{\sqrt 2}
	\end{align}
	having a GHZ-type structure of two qutrits, which yields $g_1=1/2$ and $g_2=1$.
	In case of $\boldsymbol P_{1:3:2,4}$ we decompose the state -- ordered as $|a^{(1)}\rangle\otimes|a^{(3)}\rangle\otimes|a^{(2,4)}\rangle$ -- as follows
	\begin{align}
		|\psi\rangle &
		=\frac{1}{2} |+\rangle\otimes |+\rangle\otimes |0,0\rangle
		+\frac{1}{2} |+\rangle\otimes |-\rangle\otimes |0,1\rangle
		+\frac{1}{2} |-\rangle\otimes |-\rangle\otimes |1,0\rangle
		+\frac{1}{2} |-\rangle\otimes |+\rangle\otimes |1,1\rangle,
	\end{align}
	which is in the form of~\eqref{Eq:UPBstatedec}.
	Applying the solution in~\eqref{Eq:UPBstatedecSolution}, we get $g_r=r/4$ for $r=1,2,3,4$.
	To show that these are the optimal eigenvalues, we calculate the reduced operator $\hat{\mathcal{L}}^{(\bar 3)}$,
	\begin{equation}
		\hat{\mathcal{L}}^{(\bar 3)}=\frac{1}{4}(|+\rangle\langle+| + |-\rangle\langle-|) \otimes (|+\rangle\langle+| + |-\rangle\langle-|),
	\end{equation}
	which represents a separable bipartite operator.
	Therefore, the eigenvectors of $\hat{\mathcal{L}}^{(\bar 3)}$ have the form $|a^{(1)}\rangle \otimes |a^{(2)}\rangle$.
	Accordingly, the SQE spectrum of $\hat{\mathcal{L}}^{(\bar 3)}$ can be split into $n$ parts through the purification by a set of $n$ orthonormal vectors.
	In this case, this set is given by $\{|0,0\rangle,|0,1\rangle,|1,0\rangle,|1,1\rangle\}$.
	For the $3$-partition $\boldsymbol P_{3:4:1,2}$ the SQE-$1$ eigenvalue can be simply obtained using the cascade structure given in Sec.~\ref{CasStr}, as $g_1=1/4$.
	Also for SQE-$2$, $3$ and $4$, we may simply apply the structure of Eq.~\eqref{Eq:PartSepRedOrth} to get the eigenvalues of $g_r=r/4$ for $r=2,3,4$.
	For the $4$-partition $\boldsymbol P_{1:2:3:4}$, the reduced operator with respect to the first party, 
	$\hat{\mathcal{L}}^{(\bar 1)}$, is of the form given in Eq.~\eqref{Eq:PartSepRed} leading to the same values for $g_r$.

 
\end{widetext}


\end{document}